\documentclass[sigconf, nonacm, 10pt]{acmart}

\makeatletter
\def\@ACM@checkaffil{
    \if@ACM@instpresent\else
    \ClassWarningNoLine{\@classname}{No institution present for an affiliation}%
    \fi
    \if@ACM@citypresent\else
    \ClassWarningNoLine{\@classname}{No city present for an affiliation}%
    \fi
    \if@ACM@countrypresent\else
        \ClassWarningNoLine{\@classname}{No country present for an affiliation}%
    \fi
}
\makeatother

\usepackage{tikz}
\usepackage{amsmath}

\usepackage{enumerate} 
\usepackage{listings}
\usepackage[obeyFinal,textsize=tiny]{todonotes}
\usepackage{verbatim}
\usepackage{paralist}
\usepackage{xspace}
\usepackage{multirow}
\usepackage{subcaption}
\usepackage{booktabs}
\usepackage{adjustbox}
\usepackage[nameinlink,noabbrev]{cleveref}

\graphicspath{{graphics/}}

\definecolor{darkblue}{rgb}{0,0,0.4}
\definecolor{darkpurple}{rgb}{0.4,0,0.4}
\definecolor{darkred}{rgb}{0.8,0,0}
\definecolor{commentgreen}{rgb}{0.1,0.4,0.1}
\definecolor{stringcolor}{rgb}{0.1,0.1,0.7}

\lstdefinestyle{python}{
    language=Python,
    morekeywords={with}
}

\lstdefinestyle{C++}{
    language=C++
}

\AtBeginDocument{%
  }

\setcopyright{acmcopyright}
 \copyrightyear{2025} 
\acmYear{2025} 
\setcopyright{cc}
 \setcctype{by}
 \acmConference[SoCC '25]{ACM Symposium on Cloud Computing}{November
 19--21, 2025}{Online, USA}
 \acmBooktitle{ACM Symposium on Cloud Computing (SoCC '25), November
 19--21, 2025, Online, USA}\acmDOI{10.1145/3772052.3772265}
 \acmISBN{979-8-4007-2276-9/2025/11}

\acmSubmissionID{pap369}

\usepackage{xspace}
\usepackage{xcolor} 
\usepackage[normalem]{ulem}

\newcommand{\VLCs}{VLCs\xspace}
\newcommand{\VLC}{VLC\xspace}
\newcommand{\Monitor}{VLC Monitor\xspace}
\newcommand{\VLCShim}{Service VLC\xspace}
\newcommand{\libtorch}{LibTorch\xspace}
\newcommand{\pytorch}{PyTorch\xspace}
\newcommand{\openmp}{OpenMP\xspace}
\newcommand{\openblas}{OpenBLAS\xspace}
\newcommand{\VLCXX}{VLC++\xspace}
\newcommand{\PyVLC}{PyVLC\xspace}
\newcommand{\VLCsfullname}{Virtual Library Contexts\xspace}
\newcommand{\LLVfullname}{library-level virtualization\xspace}

\newboolean{anonymize}
\setboolean{anonymize}{true}
\newboolean{publicversion}
\setboolean{publicversion}{false}
\newboolean{squeeze}
\setboolean{squeeze}{false}
\newboolean{desperatesqueeze}
\setboolean{desperatesqueeze}{false}
\ifthenelse{\boolean{desperatesqueeze}}{
\usepackage{titling}
\setlength{\droptitle}{-5em}   
\date{\vspace{-1em}}
\setboolean{squeeze}{true}
}{
}

\newboolean{shortbk}
\setboolean{shortbk}{false}

\ifthenelse{\boolean{squeeze}}{
\renewcommand{\baselinestretch}{1.00}
\addtolength{\belowcaptionskip}{-10pt}
\addtolength{\abovecaptionskip}{-6pt}
\setlength{\marginparsep}{15pt}
\setlength{\marginparwidth}{0.75in}
}{}

\begin{document}

\date{}

\title{\VLCs: Managing Parallelism with Virtualized Libraries}

\author{Yineng Yan$^*$, William Ruys$^*$, Hochan Lee$^*$, Ian Henriksen$^*$, Arthur Peters$^*$  \newline  Sean Stephens$^*$, Bozhi You$^*$, Henrique Fingler$^*$, Martin Burtscher$^\dagger$, Milos Gligoric$^*$ \newline Keshav Pingali$^*$, Mattan Erez$^*$, George Biros$^*$, Christopher J. Rossbach$^{*\ddagger}$}

\def \authors{\hspace{-0.2em} Yineng Yan, William Ruys, Hochan Lee, Ian Henriksen, Arthur Peters, Sean Stephens, Bozhi You, Henrique Fingler, Martin Burtscher, Milos Gligoric, Keshav Pingali, Mattan Erez, George Biros, Christopher J. Rossbach}

\affiliation{%
 \institution{\vspace{0.3em}$^*$The University of Texas at Austin, \hspace{0.3em}$^\dagger$Texas State University, \hspace{0.3em}$^\ddagger$Microsoft}
}

\renewcommand{\shortauthors}{Yan et al.}

 \begin{abstract}
As the complexity and scale of modern parallel machines continue to grow, programmers increasingly rely on composition of software libraries to encapsulate and exploit parallelism. However, many libraries are not designed with composition in mind and assume they have exclusive access to all resources. Using such libraries concurrently can result in contention and degraded performance. Prior solutions involve modifying the libraries or the OS, which is often infeasible.

We propose \emph{\VLCsfullname{} (\VLCs{})}, which are process subunits that encapsulate sets of libraries and associated resource allocations. 
\VLCs control the resource utilization of these libraries without modifying library code. This enables the user to partition resources between libraries to prevent contention, or load multiple copies of the same library to allow parallel execution of otherwise thread-unsafe code within the same process. 

In this paper, we describe and evaluate C++ and Python prototypes of \VLCs{}. Experiments show \VLCs{} enable a speedup of up to 2.85$\times$ on benchmarks including applications using OpenMP, OpenBLAS, and LibTorch. Source code of \VLCs{} is available at https://github.com/pecos/Virtual-Library-Context.

\end{abstract}



\maketitle

\section{Introduction}
Configuring and composing external libraries within the same application can be challenging. 
Many modern libraries rely on parallelism internally and make strong assumptions about the system resources their runtimes use. 
Libraries such as \openmp~\cite{openmp}, \openblas~\cite{openblas}, Qthreads~\cite{qthreads}, RaftLib~\cite{RaftLib}, TBB~\cite{tbb}, PTask~\cite{ptask}, Galois~\cite{galois}, and ARPACK~\cite{arpack}, may assume they own all system resources or only allow configuration at the process level (e.g., through environment variables). 
When these libraries are not aware of each other's resource usage, they can over-allocate resources causing degraded performance.
Even when configuration options are available, they are often limited in granularity and flexibility. Options may not be sufficient to avoid contention between libraries or to optimally exploit parallelism.

\begin{figure}[h]
\centering
\hfill
\includegraphics[width=\columnwidth]{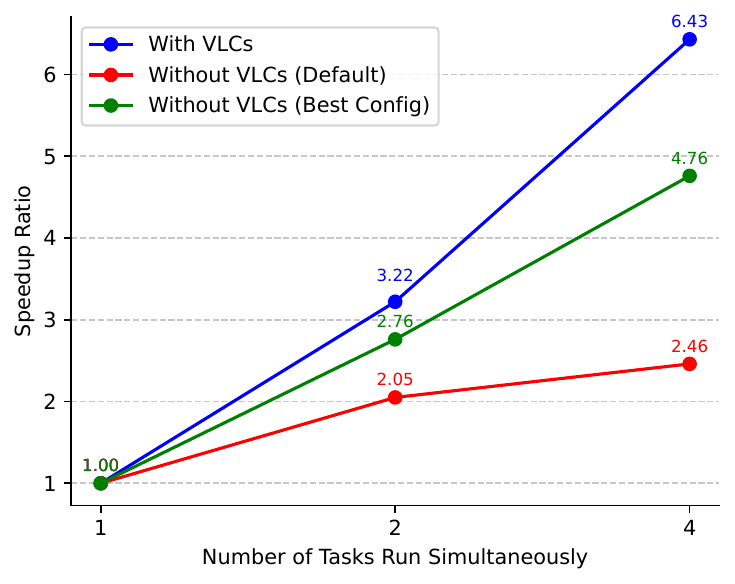}
\caption{The speedup ratio of parallel hyperparameter tuning on a Transformer model in C++ \libtorch relative to sequential hyperparameter tuning.
}
\label{fig:contention}
\end{figure}

Consider an application that uses \openblas and \openmp to orchestrate concurrent linear algebra operations, such as calling \openblas kernels from \openmp threads.
By default, both libraries spawn one thread per core, leading to oversubscription and severe resource contention. OpenBLAS can be configured at the process level, but if threads process matrices of different sizes this equal allocation may be suboptimal.
Contention can also arise from concurrent use of a single library, as has been reported for \openmp\cite{bolt}. In \autoref{sec:motivation}, we demonstrate that this slowdown, compared to an optimal allocation of resources, can be significant. Library APIs do not always allow the flexibility to effectively partition resources among threads. 

Current solutions for managing cross-library interference require modifications to library code or the OS kernel. 
This limits their practical usage and adoption, especially in scientific software which is often built on a stack of legacy libraries not under the developers' direct control \cite{bangerth2025experienceconvertinglargemathematical}.
Lithe~\cite{lithe} provides a standardized interface and thread abstraction primitives to coordinate resources with a shared runtime.
Parallel libraries must be modified to use its runtime API.
This limits its adoption on alternative, legacy, and closed-source runtime systems: only TBB~\cite{tbb} and \openmp{} have been ported to Lithe.
Bolt~\cite{bolt} provides an alternative \openmp{} runtime that avoids oversubscription when calling multiple \openmp-parallelized libraries with user-level threads. 
Effort is required to maintain the alternative runtime, and it does not provide a solution for non-\openmp{} based workloads.
Light-Weight Contexts (LwCs)~\cite{lwCs} address contention between libraries with separate resource and protection domains within a single process. 
However, this isolation requires kernel level modifications, limiting usage on managed computing clusters.

We introduce \emph{\VLCsfullname} (\VLCs), an easy-to-use tool for controlling library composition that does not require modifying library code. \VLCs are a user space, lightweight, library virtualization layer that enables users to load distinct sets of libraries into a single process and associate each set with specific system resources (e.g., a set of cores or GPUs). Different libraries can be allocated disjoint resources to mitigate contention. 
When a library is loaded into a \VLC{}, its resource-management API calls are interposed and virtualized by intercepting system calls and file system accesses used to query system resources. \VLCs require no modification of the library code and no recompilation of the library binary.
For example, the available CPU cores visible to libraries within one \VLC{} can be limited such that they will not be oversubscribed when used with libraries in other \VLCs{}.

\VLCs{} provide performance isolation but not data isolation.
Libraries between \VLCs{} still efficiently share data in the same address space. 
By loading multiple instances of the same library into separate \VLCs{}, the instances can run in parallel using distinct internal static state and separate partitions of resources.
This enables applications to make safe simultaneous calls into libraries that are otherwise not thread-safe. While \VLCs enable application control over resources, finding the right resource configuration is non-trivial. \VLCs solve this with an auto-tuner that automatically searches the configuration-space. This helps the user identify (often counterintuitive) optimal configurations.

We implement \VLCs for C++ and Python and present several use cases based on a set of common and important libraries, including \openmp, \openblas, \libtorch~\cite{pytorch}, Kokkos~\cite{kokkos}, and ARPACK~\cite{arpack}. We show that \VLCs{} enable efficient composition of libraries within a single-process application; current alternatives require the use of multiple processes and explicit inter-process communication. These examples overcome composability challenges that either degrade performance or prevent the application from running correctly, including those relating to libraries that either: (1) assume they own all process resources; (2) can utilize only a subset of available resources (e.g., a single GPU or a single core); or (3) assume a static and inflexible resource allocation. Our experiments show that the \VLC mechanism has low overhead, allowing \VLCs to increase performance for applications with multiple concurrent libraries that contend. \VLCs achieve a speedup of up to $2.85\times$ on benchmarks, a $1.41\times$ speedup on a multi-GPU Heat3D application using Kokkos, and a $1.96\times$ speedup on the ARPACK eigenvalue solver.
We make the following contributions:

\begin{itemize}
    \item We introduce the novel concept of library-level virtualization that offers fine-grained resource management without requiring library or OS modifications. Programmers can adopt \VLCs with as few as 10-20 lines of application code changes.

    \item We demonstrate how library-level virtualization helps avoid resource contention, improves nested parallelism, and enables parallelization of thread-unsafe calls.

    \item We implement C++ and Python prototypes to demonstrate the applicability of \VLCs on compiled and interpreted languages.

\end{itemize}

\begin{figure}[h]

\includegraphics[width=1.00\columnwidth]{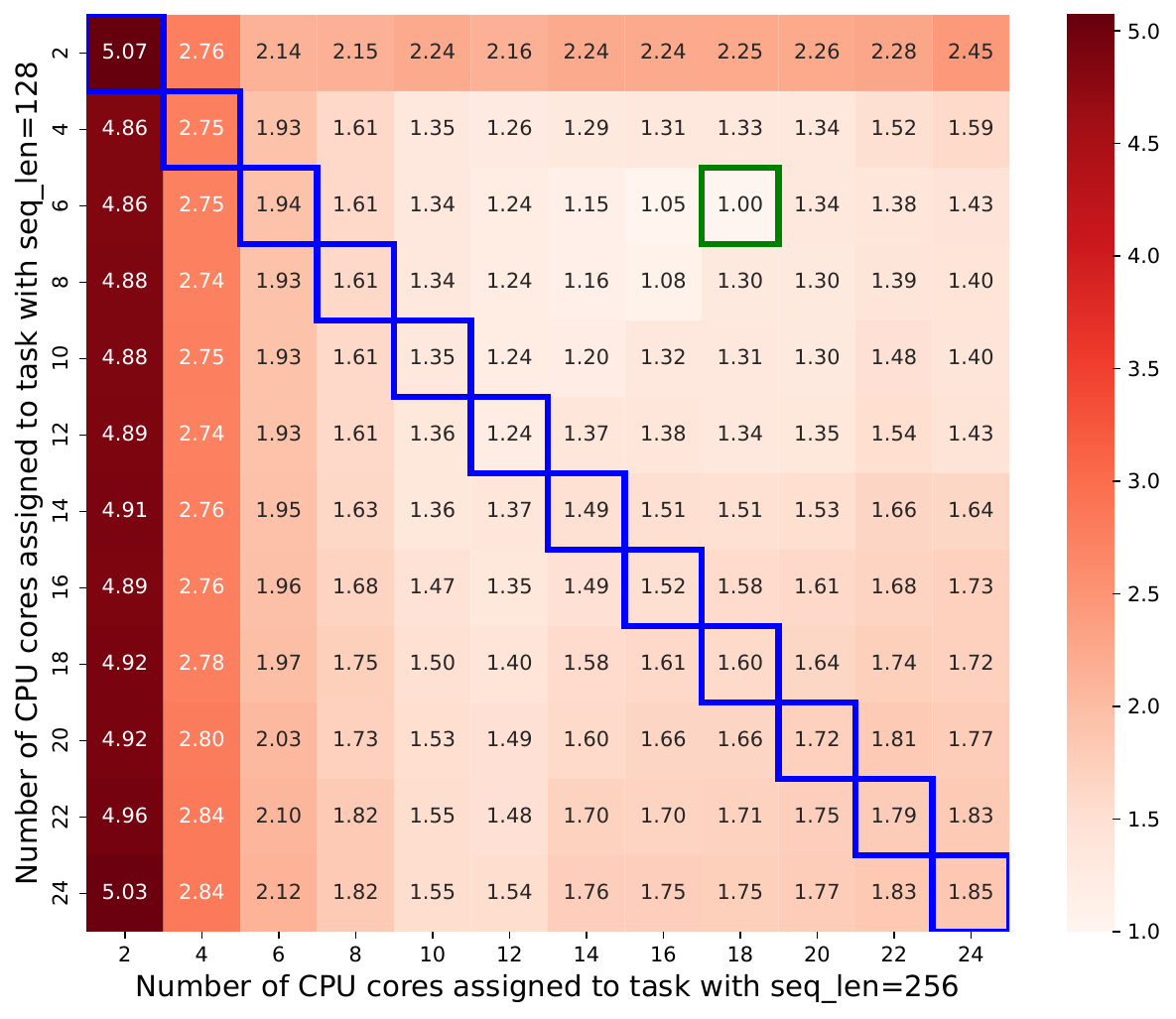}
\caption{Heatmap of relative execution times for CPU core partitions between two concurrent hyperparameter tuning tasks. Lighter regions indicate shorter execution times. The optimal partition is marked with a green box. Blue boxes represent partitions achievable with \libtorch APIs. Without \VLCs, the optimal partition cannot be achieved.}
\label{fig:heatmap}
\end{figure}

\section{Motivation}
\label{sec:motivation}
Complex application codes often combine multi-threaded libraries that do not always support fine-grained management of resources. 

{\tt FFTW}~\cite{FFTW}, {\tt SuperLU}~\cite{superlu}, {\tt OpenCV}~\cite{opencv}, and {\tt Sundials}~\cite{sundials} 
allow setting the number of threads globally, but do not allow allocation of specific threads to a call. The latter may result in suboptimal memory access and unnecessary communication overheads. {\tt Mumps}~\cite{MUMPS} supports an arbitrary number of threads but only through the environment variable {\tt OMP\_NUM\_THREADS} which affects not just {\tt Mumps}, but libraries that rely on \openmp.

Many numerical methods require nested multi-library calls. Examples include:  multigrid methods involving relaxation, restriction, and prolongation steps that at different levels may require a different number of threads \cite{multigrid}; Hierarchical matrix and N-body codes in which different phases may require a different number of resources for optimal performance \cite{nbody2}; Multiphysics and adaptive mesh refinement methods that may require radiation solvers, stencil solves, semi-Lagrangian steps, particle-in-cell steps, or boundary and internal interface physics that again require fine-grained control for optimal performance \cite{amrex}.  Furthermore, tasks like optimization, uncertainty quantification, and machine learning induce further diversity of libraries and workloads. 
Controlling resources in hierarchical and nested scientific computing modules can be quite challenging. 

\openmp spawns as many threads as there are logical cores by default. When an application composes \openmp 
with other parallel libraries, each library may allocate a thread pool equal to the number of cores. This can lead to oversubscription, where threads compete for a limited number of cores. Developers have reported issues with degraded performance when combining \openmp with parallel libraries like \openblas~\cite{openblas_issues1}. Additionally, a persistent problem arises when two libraries using \openmp are composed: they cannot set different \openmp thread-pool sizes because they share the same \openmp instance and configuration. This issue remains unresolved for \pytorch, Numba~\cite{numba-issue}, Ray~\cite{ray-issue}, and scikit-learn~\cite{sklearn-issue}.
Similarly, while it is possible to manually set the number of available threads globally, libraries like \pytorch (and C++ \libtorch) do not provide an API to configure thread allocation at a per-operation granularity. 

{\textbf{Example}}
Hyperparameter tuning workflows for machine learning models can be accelerated by parallelizing runs to leverage the large memory capacity and massive parallelism \cite{hyperparameter,hpc_hyperparameter}. When running on the same node, it can be beneficial to run within a single process to efficiently share large datasets and intermediate computations \cite{hippo}. We implement a C++ \libtorch parallel hyperparameter tuning workflow on a transformer-based language model with 8 heads, 6 layers, and a 512 embedding size. The training dataset is {\tt wikitext2} \cite{wikitext}.

We measure total training time for tuning a fixed number of hyperparameters with different numbers of concurrent tasks. \autoref{fig:contention} shows the speedup relative to sequential tuning on a two-socket ARM architecture supercomputer node with a 144-core NVIDIA Grace CPU Superchip and 237 GB of DRAM. In the default \libtorch configuration, each of the training tasks spawns 144 threads (matching the total number of cores). Running 4 tasks concurrently creates 576 threads, severely oversubscribing the 144 cores and leading to poor scaling due to contention.

Without fine-grained control, such as that provided by \VLCs, the best alternative using the current PyTorch API is to globally limit the threads per task (e.g., 36 threads each for 4 tasks). This prevents CPU oversubscription but results in suboptimal resource utilization, as models with different hyperparameters show varying scalability and execution times. Assigning them an equal number of threads fails to maximize parallelism. \VLCs enable each task to be allocated a disjoint set of cores with a tailored number of threads. This strategy eliminates contention while improving core utilization. As shown in \autoref{fig:contention}, \VLCs achieve up to a 6.43$\times$ speedup over the sequential baseline and 2.61$\times$ over the default concurrent configuration.  \VLCs outperform the best achievable configuration with the current PyTorch API by 1.35$\times$.

\autoref{fig:heatmap} illustrates how execution time varies when \VLCs assign different cores to two tasks. This experiment specifically involves tuning a model with different sequence lengths (128 and 256) simultaneously on a 24-core x86 computer. The best performance is achieved by allocating 6 cores to the task with 128 sequence length and 18 cores to the 256 sequence length task. Since \libtorch does not support assigning different numbers of threads to each task, its only tuning option without \VLCs is represented by the diagonal of the heatmap (highlighted with blue boxes). Without \VLCs, it fails to achieve the best performance as the optimal configuration is not on the heatmap diagonal.

\section{Related Work}

\begin{figure}[!b]
\centering
\includegraphics[width=\columnwidth]{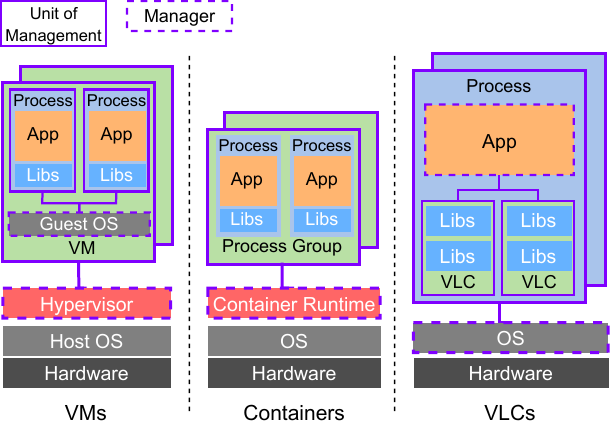}
\caption{Overview of \VLCs and other virtualization techniques. Dashed boxes represent resource managers and solid boxes are the unit of management.}
\label{fig:comparison}
\end{figure}

\paragraph{\textbf{Composition of Libraries}}
Software design techniques that separate work from workers can, in principle, solve the problem that we address with \VLCs. For example, Lithe~\cite{lithe} enables efficient library composition by requiring libraries to expose APIs to an external thread manager.
This enables performance improvements in applications with mixed or nested parallelism. Lithe requires the libraries to be ported to use those APIs. The adoption of such techniques has not been widespread.
To the best of our knowledge, only TBB and \openmp have adopted Lithe techniques.

\VLCs, in contrast, require no changes to existing libraries.
Like \VLCs, libcapsule \cite{libcapsule} enables the use of incompatible libraries in the same application. However, it provides no mechanisms for resource management.
For applications composed of multiple libraries that are all parallelized by \openmp, Bolt~\cite{bolt} is a new runtime design using user-level threads to reach a balance between minimizing thread oversubscription while maximizing parallelism. However, this approach is \openmp specific and requires a new implementation of the \openmp library. A more general, non-library-specific solution, such as \VLCs, is desirable.

\paragraph{\textbf{Virtualization}} Containers \cite{docker, x-containers} and VMs provide a way to isolate runtime environments, but they provide no support for simple data sharing between components. In fact, they are explicitly designed to prevent this kind of direct communication, which renders them inappropriate for separating components of a single application. 
\VLCs support the virtualization of resources at library granularity within a single address space where data are shared natively. Figure \ref{fig:comparison} illustrates the layers of resource management in different virtualization techniques. \VLCs provide a new technique for resource virtualization, which allows applications to manage the resources available to libraries.

Library OSes \cite{lib_os_1, lib_os_2, lib_os_3, lib_os_4}, Unikernels \cite{unikernel_1, unikernel_2, unikernel_3, unikernel_4, unikernel_5, unikernel_6}, and lightweight virtualization techniques such as Xax \cite{xax} and picoprocesses \cite{picoprocess} have the goal of giving applications more granular and flexible control over low-level resources. These techniques require special kernel support or emulators, which makes the application non-portable.

OS-level mechanisms including \texttt{cgroups} and \texttt{cpuset} provide a way to limit the resource usage at the process level. In contrast, \VLCs provide a  fine-grained resource isolation and management at an intra-process level.

\paragraph{\textbf{Subunits of Processes}}
Process-in-Process (PiP) \cite{pip} loads entire programs into a shared address space, typically using an optimized MPI runtime for communication. In contrast, \VLCs focus on managing libraries within a single process. Communication between \VLCs occurs directly via function calls and standard shared memory mechanisms without MPI.

Lightweight isolation techniques such as LwCs \cite{lwCs}, CubicleOS \cite{cubicle}, and Mondrix \cite{mondrix} provide sub-address space \emph{data} isolation. These systems require source-level changes and are motivated by increasing isolation for security. Additionally, LwCs require special kernel support. \VLCs require no source code modification or recompilation of existing libraries and are motivated by providing performance isolation. 

\section{Design}
\VLCs are performance-isolated execution environments that encapsulate sets of libraries in a single process. Each \VLC can have different assigned  resources and environment variables.
With \VLCs, users can force a library to only use specific resources even if the library does not provide resource management APIs.    

Libraries do not have to be unique to a \VLC; the same library may be instantiated in several \VLCs with different configurations in each. 
Each \VLC has its own linker namespace and static state, allowing multiple, even potentially incompatible, library versions to coexist in one process. 

\begin{figure}[h]
\centering
\includegraphics[width=0.8\columnwidth]{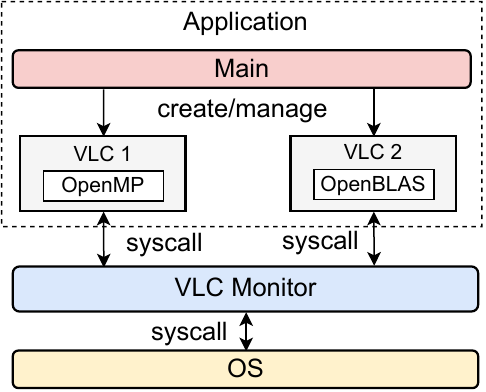}
\caption{Programming model of \VLCs. \openmp and \openblas are loaded into separate \VLCs with different resource allocations; Resource query system calls are interposed by the \Monitor.}
\label{fig:vlc-arch}
\end{figure}

\autoref{fig:vlc-arch}  illustrates the VLC system.
The application creates \VLCs via API calls to isolate libraries and assign resources. At runtime, a \Monitor transparently intercepts library system calls and virtualizes resource-related queries.

\subsection{Library Isolation}
\VLCs use \emph{linker namespaces} to provide isolation between libraries. Linker namespaces give users the ability to load a dynamic shared object (DSO) into an explicit namespace, such that libraries in one linker namespace cannot access symbols in other namespaces. This isolation prevents symbol conflicts when multiple copies of the same library are loaded into a single process. Since all libraries are still in the same address space, data can be shared between them through virtual memory. Multiple instances of the same DSO can be loaded into different namespaces without conflict by giving each instance distinct static internal state. This enables parallel calls into a library that may not be thread-safe since a call that changes the state of one copy of the library will not affect the state of the other. This also enables loading potentially incompatible versions of a library into the same process, similar to libcapsule~\cite{libcapsule}. This feature is discussed in detail in \autoref{sec:incompatible}.

\subsection{Resource Virtualization}
The \VLC runtime virtualizes system resources such as the number of CPU cores available to a \VLC via system call interposition and virtualized resource configuration files. When an application initializes the \VLC runtime, a \Monitor process is forked to interpose system calls from the application process. We use \texttt{ptrace} to intercept system calls because it operates entirely in user space. This allows our tool to be used on externally-managed systems, like supercomputers, where users lack root privileges, unlike kernel-module-based tools such as SystemTap~\cite{systemtap}. To reduce interception overhead, \texttt{Seccomp BPF} is used as a filter to avoid intercepting system calls that are not related to resource management.

Resource-query system calls are interposed by the \Monitor process. When libraries within a \VLC attempt to query available system resources, the responses are modified so that the caller perceives only the resources allocated to the \VLC. When a resource query system call is invoked multiple times during a library’s life cycle, the forged resource result may be cached for each \VLC to reduce overhead. Some libraries query resources by accessing resource configuration files directly. Galois~\cite{galois}, for example, counts the number of CPU cores by reading \texttt{/proc/cpu} files. Resource visibility for these libraries is restricted by redirecting I/O requests to virtual resource configuration files that list only the resources the user has assigned to the \VLC. To ensure threads within a \VLC are scheduled exclusively on their allocated cores, the \Monitor sets CPU affinity for all threads. Additionally, any \texttt{sched\_setaffinity} system calls made by libraries in a \VLC are interposed to prevent them from scheduling threads on cores outside their assigned set.

\subsection{\VLC Configuration and Management}

\VLCs are assigned resources by the application programmer or the \VLCs auto-tuner.
The assignments of different \VLCs may overlap if desired.
Resource allocations can be configured before a library is loaded and can be adjusted at any time. However, libraries and their associated thread pools within a \VLC may not adapt well to dynamic changes. Environment variables for a specific \VLC can also be reconfigured during program execution, though the frequency with which a library checks these variables is beyond our control. An alternative approach is to create multiple \VLCs of the same library with different configurations and switch between them as needed. The low overhead of entering and exiting a \VLC (see \autoref{microbenchmarks}) makes this approach practical.

\begin{figure}[!t]
\includegraphics[width=\columnwidth]{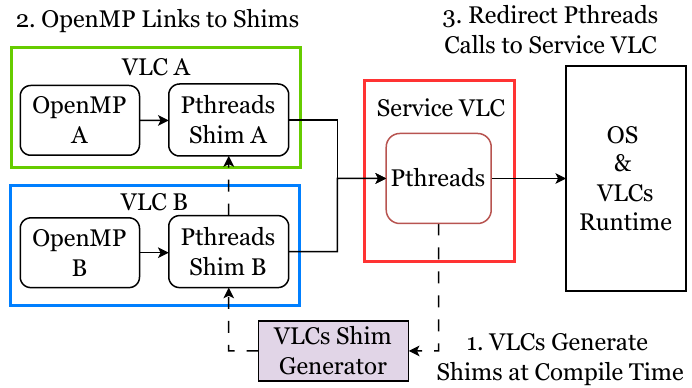}
\caption{\VLCShim Control Flow. \openmp is loaded into a \VLC and links to the generated shim of pthreads. The calls to pthreads are redirected to Service \VLC.}
\label{fig:vlc-service}
\end{figure}

\subsection{\VLCShim}
\label{sec:vlcshim}
In Linux, a linker namespace is created by calling \texttt{dlmopen}. However, a few libraries including TBB, CUDA, and older versions of pthreads (prior to glibc 2.34)~\cite{pthread_bug,tbb-bug} are not compatible with \texttt{dlmopen}.
A common cause for this incompatibility is the library calling \texttt{dlopen} internally. A nested call to \texttt{dlopen} in a linker namespace created by \texttt{dlmopen} causes a segfault due to a bug in \texttt{dlmopen} \cite{dlmopen-bug}. 
To address these incompatible cases, we provide a \VLCShim to load shared instances of the problematic libraries. 

The \VLCShim is a special context into which a \texttt{dlmopen}-incompatible library is loaded globally via \texttt{dlopen}.
Unlike \texttt{dlmopen}, which creates a new linker namespace, \texttt{dlopen} loads libraries into the base namespace and avoids creating another copy of the library if it is already loaded. 
We provide a script to automatically generate shims for libraries loaded into the \VLCShim. 
These shims replicate library symbols for each context, making them available to other \VLCs. 
When libraries in a \VLC access methods from a \texttt{dlmopen}-incompatible library, the shim performs API forwarding to the underlying globally loaded library in the \VLCShim.
This ensures libraries in \VLCs can depend on libraries that are incompatible with \texttt{dlmopen}.

\autoref{fig:vlc-service} illustrates how the \VLCShim allows \openmp to utilize the \texttt{dlmopen}-incompatible pthreads library within a \VLC. Before a \VLC is created, pthreads is loaded into the \VLCShim. 
The addresses of pthreads methods are dumped into a page shared with the shim to direct where calls should be forwarded. When \openmp is loaded into the \VLC, a pthreads shim is loaded as its dependency. The shim is generated by the \VLCs Shim Generator at compile time. When \openmp is running, the shim is responsible for redirecting all pthreads calls to the \VLCShim. Exchanging function addresses via a shared page occurs only once during \VLCShim initialization. These addresses are cached in a jump table to minimize the cost of forwarding API calls.

The \VLCShim is transparent to libraries that depend on it. It does not require any code modifications or recompilation of these libraries or applications. The overhead of forwarding an API call through the shim is negligible (see \autoref{microbenchmarks}), as it is implemented in just 23 lines of assembly code and incurs a constant cost regardless of the workload.

\subsection{Programming Interface}
\label{sec:interface}

We provide an API in C++ and Python to create \VLCs and manage their resources, listed in \autoref{tab:vlc-api}. \autoref{fig:python-interface} shows an example of using the Python interface for \VLCs.
The \VLCs \lstinline|a| and \lstinline|b| are created and configured in Lines~\ref{line:vlcs}--\ref{line:end-config}.
Each isolated execution environment may configure which cores are visible to libraries within that environment, as in Line~\ref{line:end-config}. Control flow can enter and exit a \VLC, Line~\autoref{line:begin-computation}, using Python's context manager \lstinline{with} syntax.
    
\autoref{fig:cpp-interface} shows the same example using the C++ interface. When the application starts, it initializes the \Monitor on Lines~\ref{line:monitor-create}--\ref{line:monitor}. For illustration purposes, we create two parallel threads using standard C++ on Lines~\ref{line:thread-a} and \ref{line:thread-b}. 
On Lines~\ref{line:vlcs-create-a} and \ref{line:vlcs-create-b} two \VLCs are created with the \VLC ID and thread ID as arguments. The visible cores need to be configured for each \VLC before the libraries are loaded. To load a library into a \VLC, a \texttt{VLC::Loader} object is created with the library path and a pointer to an associated \texttt{VLC::Context}.
The last boolean argument indicates whether transparent mode is enabled (see Section \ref{sec:transparency}). By running in transparent mode, there is no need to load function pointers from namespaces manually. Once the library is loaded into a \VLC, its methods become available for use. This minimizes the code changes required when using \VLCs.

\begin{figure}[h]
\begin{lstlisting}[style=python,columns=fullflexible]
a, b = VLC(), VLC()$\label{line:vlcs}$
a.set_allowed_cpus([0])
b.set_allowed_cpus([1,2,3,4,5,6,7])$\label{line:end-config}$
with a:$\label{line:begin-computation}$
    import numpy 
    async_computation()
with b: 
    import numpy 
    async_computation()$\label{line:end-computation}$
\end{lstlisting}
\caption{\label{fig:python-interface}
An example of \VLCs usage in Python.
}
\end{figure}
\begin{figure}[h]
\begin{lstlisting}[style=C++,columns=fullflexible]
int main(int argc, char ** argv) {
    VLC::Runtime monitor; $\label{line:monitor-create}$
    monitor.initialize(); $\label{line:monitor}$
    std::async([&]() { /* on thread 1 */ $\label{line:thread-a}$
        VLC::Context a(1, gettid());$\label{line:vlcs-create-a}$
        a.set_allowed_cpus("0-11");$\label{line:vlcs-config-a}$
        VLC::Loader loader(&a, "lib.so", true);
        computation();
    });
    std::async([&]() { /* on thread 2 */ $\label{line:thread-b}$
        VLC::Context b(2, gettid());$\label{line:vlcs-create-b}$
        b.set_allowed_cpus("12-23");$\label{line:vlcs-config-b}$
        VLC::Loader loader(&b, "lib.so", true);
        computation();
    });
}
\end{lstlisting}
\caption{\label{fig:cpp-interface}
An example of \VLCs usage in C++. \lstinline{std::async} is used to simplify thread allocation.} 
\end{figure}

\begin{small}
\begin{table}[bth]
\centering
\vspace{.25cm}
\begin{tabular}{l|p{2in}}
\lstinline|VLC::Monitor (C++)| & Create the \Monitor \\
\lstinline|initialize (C++)| & Initialize the \Monitor \\
\lstinline|VLC::Loader (C++)| & Open a shared library and \\ & its dependencies in this \VLC \\
\lstinline|VLC() (Python)| & Create a \VLC \\
\lstinline|__enter__ (Python)| & Mark  \VLC as the default \\ & for imports in this thread \\
\lstinline|__exit__ (Python)| & Restore the state of imports \\ & to what it was before  \\ & \lstinline|__enter__| was called\\
\lstinline|set_allowed_cpus (Both)| & Make only  specific set of \\ & CPUs visible to this \VLC  \\
\lstinline|setenv (Both)| & Set an environment variable \\ & in  this \VLC \\
\lstinline|unsetenv (Both)| & Unset an environment  \\ & variable in this \VLC \\
\end{tabular}
\caption{\VLC API. \lstinline{setenv}, \lstinline{unsetenv}, and \lstinline{set_allowed_cpus} have the same interface in C++ and Python. The \lstinline{__enter__} and \lstinline{__exit__} routines are called implicitly in Python's \lstinline[style=python]{with} syntax. \label{tab:vlc-api}}\vspace{-5pt}

\end{table}
\end{small}

\subsection{Programmability and Transparency}
\label{sec:transparency}

\VLCs are designed to allow existing applications to use them with minimal code changes. To achieve this goal, the use of \VLCs should be transparent so most of the application code can remain intact. The only new code that the developer must write is to configure and create a \VLC.

One limitation in C++ is that symbols from libraries in the newly created linker namespace are not automatically visible to other namespaces. To call functions from a library loaded by \texttt{dlmopen}, one must obtain function pointers by calling \texttt{dlsym} explicitly. This must be done for every function and, without tooling, could become a burden on the developer when the number of functions is large. \VLCs avoid the need to use \texttt{dlsym} explicitly on every function by providing a ``transparent mode'' in the C++ implementation. This mode uses automatically generated library shims that redirect calls through a runtime jump table (see \autoref{sec:vlcxx}). We achieve similar transparency in Python by overriding \texttt{builtin.\_\_import\_\_} and \texttt{getattr} (see \autoref{sec:pyvlc}).

\section{Implementation}
\label{sec:implementation}

We prototyped \VLCs in C/C++ and Python.
Due to differences in how shared objects are managed and how they load dependencies, there are differences in the details of each implementation. 

\subsection{\VLCXX: \VLCs for C/C++}
\label{sec:vlcxx}

\VLCXX is a prototype of \VLCs for C/C++ applications. \VLCs provide namespaces and resource management contexts, while the \Monitor provides a supervisor for lifecycle management APIs and system call interposition. When libraries allocate resources, they often must first query the system to learn what system resources are available.

\VLCXX creates an interposition layer between libraries and the OS to virtualize the visible resources. Whenever a library makes a system call, it is inspected by the \Monitor using \texttt{ptrace} and potentially modified to reflect the resources assigned to the current \VLC.

Interposing every system call is both expensive and unnecessary. Most libraries will not query system resources frequently, and most of the system calls made by the application are unrelated to \VLCs. \VLCXX uses \texttt{Seccomp BPF} to reduce overhead and filter system calls before interposition by the monitor. 
Only targeted system calls like \texttt{sched\_getaffinity} are intercepted by \texttt{ptrace} in the \VLCs Monitor process. 
\texttt{Seccomp BPF} can filter out most of the system calls without triggering a more expensive trap to \texttt{ptrace}. To demonstrate that this overhead is minimal, experiments in \autoref{table:programmability} show the overhead introduced by \VLCs themselves is less than 0.54\%.

To support libraries that query system resources through \texttt{/proc/} files, \VLCXX provides a virtualized resource filesystem. When requesting files such as \texttt{/proc/cpuinfo}, the $openat$ system call will be interposed. This checks the filename in the system call arguments and, if the filename matches a resource file, it will be modified and redirected to a forged resource file. 
The forged resource file is dynamically generated during \Monitor initialization and only shows the subset of resources assigned to the \VLC.

\paragraph{Transparency.} \VLCs maintain transparency and avoid explicit use of function pointers at the source level. 
In transparent mode, a library shim is linked to the application in place of the target library that is loaded into a \VLC. 
When a \VLC is created, the target library is loaded by \texttt{dlmopen} and its function addresses are resolved by the \VLCXX runtime using \texttt{dlsym}. The \VLCXX runtime then saves the function address in a jump table maintained by the shim. For each function call, the application calls the shim it is linked to. Then, the shim dispatches the call following the jump table so the call can be executed by a copy of the target library in the \VLC. To distinguish which \VLC the call should be executed in, the \VLCXX runtime and the shim maintain a mapping between thread id and \VLC id. All these steps are transparent to the application. 

\subsection{\PyVLC: \VLCs on Python}
\label{sec:pyvlc}
Typical C++ applications load dynamic libraries and their shared object dependencies using the dynamic linker when the application is loaded.
However, modules in Python applications are loaded as the application is interpreted. 
Supporting these differences in how shared libraries are loaded, as well as managing the interface to forward methods from Python modules into specific \VLC contexts requires the development of a separate \VLC prototype for Python. 

For simplicity, \PyVLC prototype uses the simpler approach of interposing glibc calls for resource management, though it is straightforward to extend this to the more powerful system call interception approach of \VLCXX.

The Python import mechanism can be customized by overriding the \texttt{builtins.\_\_import\_\_} function without modifying the interpreter.
This enables redirecting any module import to the appropriate \VLC. To prevent unnecessary duplication of certain modules, they can be marked as \emph{exempt} from the special import handling at the Python level. Such modules are resolved using the default import mechanism into the global namespace. This provides similar functionality to the Service \VLC used in the C++ implementation.
As the Python interpreter is shared across \VLCs, we exempt modules like \lstinline{sys} that are directly connected to its core capabilities. We also exempt standard library modules by default.

When a module is loaded into a new namespace, its symbols are not available for resolution of subsequently loaded libraries \cite{dlmopen}. This prevents a module from accessing its dependencies even if all of them are loaded in the same linker namespace. 
To lift this limitation, we apply a 31-line code patch \cite{glibc-global-patch} to the dynamic linker $libdl$ that enables support for the \texttt{RTLD\_GLOBAL} flag for \lstinline{dlmopen}. 
In contrast, our C++ prototype does not require this patch as a library and its dependencies are loaded all at once by the dynamic linker, instead of dynamically as needed. Symbols of dependencies have been resolved as soon as this loading is finished. 

Several internal module caches must be overridden to ensure state isolation in Python \VLCs:

{\textbf{Shared Object-Handles Cache:}}
All current Python versions (3.14 and older) maintain a
finite cache of shared object handles. Entries of this cache are
reused to avoid calls to \lstinline{dlopen} when possible. 
Since \VLCs are designed to allow potentially loading \emph{the exact same
  shared object} multiple times into different \VLCs, this caching
must be disabled. We disable the cache to ensure that the Python
interpreter always receives the handle from the current \VLC linker
namespace. 

{\textbf{Module-Specification Cache:}}
The Python interpreter maintains an internal cache of module specification objects associated with each native Python extension module. Prior to loading any new portion of a library, we swap all known module specification objects associated with that library from the current \VLC into the module specification cache.

{\textbf{Module Object Cache:}}
The \VLC machinery must manage the module object cache in $sys.modules$. We manage this cache by ensuring that it is correctly populated with references to objects that are appropriate for the current \VLC. We refer to the objects we insert into the cache as forwarding modules. When an import occurs that will load a new module, all forwarding modules in $sys.modules$ associated with the corresponding library are replaced by those modules in the \VLC.

These modifications are implemented by dynamically overriding the interpreter's behavior, enabling \PyVLC to operate with the native Python interpreter and ensuring its compatibility. None of these changes require modifying the Python interpreter.

\section{Evaluation}

\VLCs demonstrate low overhead, broad applicability, and effectiveness in improving performance in a range of scenarios.
All experiments are performed on a machine with two 12-core Intel E5-2650 v4 sockets, 128 GB of RAM, and four NVIDIA Tesla P100-SXM2 GPUs,
running Ubuntu 22.04 with kernel 5.15.0. \VLCXX uses the unmodified dynamic linker from glibc 2.35 and is compiled with g++ 11.4.0. 
\PyVLC is implemented on top of Python 3.9 and the patched dynamic linker from glibc 2.30. For libraries involved in the evaluation, we used Open MPI 5.0.3, \openblas 0.3.20, CUDA version 11.7, and Kokkos version 4.6.1

We conduct experiments using micro- and macro-benchmarks to address the following research questions (RQs): 
\begin{compactitem}
\item \textbf{(RQ1)} What is the overhead of using \VLCs? 
\item \textbf{(RQ2)} How much effort is required to use \VLCs? 
\item \textbf{(RQ3)} Does performance improve with \VLCs? 
\item \textbf{(RQ4)} Do \VLCs improve nested parallelism?
\item \textbf{(RQ5)} Can \VLCs parallelize thread-unsafe libraries?
\end{compactitem}

\begin{table}
    \centering 
    \begin{small}
    \begin{tabular}{l|c}
        \toprule
        Initialize \PyVLC Modules   & 198 ms \\
        Initialize \VLCXX Monitor   & 0.54 ms \\ 
        \midrule
        \VLCXX Create \VLC            & 40.2 $\mu$s   \\
        \PyVLC Create \VLC            & 3.96 ms   \\
        Enter \VLC              & 7.8 $\mu$s \\
        Leave \VLC               & 6.9 $\mu$s   \\
        Enter \VLC (w/ Affinity) & 40 $\mu$s  \\
        \midrule
        sched\_getaffinity (w/ \VLC)   & 20.44 $\mu$s  \\
        sched\_getaffinity (w/o \VLC)   & 0.64 $\mu$s  \\
        open("/proc/cpuinfo") (w/ \VLC)  & 266.59 $\mu$s \\
        open("/proc/cpuinfo") (w/o \VLC)  & 5.81 $\mu$s \\
        \midrule
        mmap (w/ \VLC)  & 0.82 $\mu$s \\
        mmap (w/o \VLC)  & 0.78 $\mu$s \\
        \midrule
        cudaMemcpy (w/ \VLCShim)  & 2.63 $\mu$s \\
        cudaMemcpy (w/o \VLCShim)  & 2.58 $\mu$s \\
        \bottomrule
    \end{tabular}
    \end{small}
        \caption{\VLC Management Benchmarks. Times are the mean of 120,000 samples.}\vspace{-15pt}
    \label{tab:overhead}
\end{table}

\subsection{Overhead}
\label{microbenchmarks}
To evaluate \textbf{RQ1}, the overhead of \VLCs, we run a set of simple microbenchmarks that 
repeatedly exercise the \VLC API and make resource-related and other system calls. 
We report the average time for each call in \autoref{tab:overhead}. 

\PyVLC API calls take much longer than those of \VLCXX, but both are reasonable in context. 
The main reason for the longer initialization time of \PyVLC is that it must load libraries fully at 
initialization time (e.g., \texttt{librt} and pthreads), while \VLCXX defers those loads.
Calls to resource-query APIs must be interposed and do incur an overhead, although it is
quite low at only $20 \mu$s for {\tt sched\_getaffinity} and $260 \mu$s when accessing {\tt /proc/cpuinfo}. 
These calls are rare because resources are not regularly reconfigured. Importantly, the runtime overhead 
of system calls that are not interposed is very low (~$2\%$). The overhead of making the library API calls themselves is negligible.

We also evaluate the overhead of loading libraries into \VLCs, as shown in \autoref{tab:loadtime}. Library loading via \VLCXX is efficient; specifying the full library filepath can lead to faster resolution than when the path is searched without \VLCs. The results are different with \PyVLC because it overrides Python's default import machinery, disables caches, and forces additional iterations over the Python interpreter's \lstinline|module.__dict__|. The conclusion is that \VLCs have reasonable overheads during initialization and for management calls, whereas overheads from the \VLCShim and for system calls that are not interposed are negligible.

In \autoref{table:programmability}, we present the end-to-end overhead of using \VLCs in applications. All overheads are below 1\%, demonstrating that the overall impact of \VLCs on application performance is minimal.

\begin{table}
    \centering 
    \begin{tabular}{l|l|c|c}
        \toprule
        Library & Prototype & VLCs & Base\\
        \midrule
        numpy & \PyVLC & 240 ms & 181 ms \\
        scipy & \PyVLC & 333 ms & 192 ms \\
        LibTorch & \VLCXX & 524 ms & 492 ms \\
        \openblas & \VLCXX & 8 ms & 11 ms \\
        \openmp & \VLCXX & 5 ms & 3 ms \\
        Kokkos & \VLCXX & 11 ms & 6 ms \\
        Arpack & \VLCXX & 11 ms & 16 ms \\
        \bottomrule
    \end{tabular}
    \caption{\VLCs Library Load Time Comparison. We show time taken to load 4 copies of the same library into \VLCs compared to standard loading without \VLCs. 
    }\vspace{-5pt}
    \label{tab:loadtime}
\end{table}

\subsection{Programmability}

\begin{table}
\begin{adjustbox}{width=0.48\textwidth}
\centering
\begin{tabular}{ c||c|c|c  }
\hline
Library Name & Application Name & LoC Changes & \VLCs Overhead\\
\hline
\multirow{3}{*}{\openmp} & Kmeans & 27 & 0.11 \%\\ 		
& Hotspot3D & 27 & 0.35 \%\\
& CFD & 27 & 0.42 \%\\
\hline
\multirow{3}{*}{\openblas} & Cholesky & 13 & 0.54 \%\\ 		
& GEMM & 13 & 0.38 \%\\
& GESV & 13 & 0.31 \%\\
\hline
\multirow{2}{*}{\libtorch} & Transformer & 27 & 0.45 \%\\ 		
& DNN & 27 & 0.14 \%\\
\hline
\end{tabular}
\end{adjustbox}
\caption{\VLC application overhead. \emph{Lines of Code Changed} shows the number of code modifications required to use \VLCs. \emph{\VLCs Overhead} shows the performance slowdown when the application runs serially within a single \VLC.}\vspace{-5pt}
\label{table:programmability}
\end{table}

To evaluate the programmability of \VLCs, \textbf{RQ2}, we measure the lines of code and the effort to optimize \VLCs resource usage. 
\VLCs do not require library code changes, but do require initialization and interaction with client libraries.
To illustrate the ease of porting an existing application to \VLCs, \autoref{table:programmability} 
details the number of lines we needed to modify in each benchmark discussed in \autoref{sec:macrobenchmark}. None 
of the codes require more than 27 lines to be changed. 

Although \VLCs allow developers to manage resources at a fine granularity, determining the optimal
resource partition is not always straightforward. To address this challenge, we developed a tuning 
framework that performs a grid search over all possible partitions and reports the optimal configuration. 
\autoref{fig:heatmap} presents a heatmap generated by our tool, illustrating which partitions yield the 
best performance for the \libtorch application discussed in \autoref{sec:motivation}. The time required by the \VLCs auto-tuner to find the optimal configuration depends on the grid size. For instance, determining the configuration for a pair of \openblas benchmarks requires 64 runs with a grid size of 3, taking approximately 10 minutes on a 24-core machine. This time can be significantly reduced if users provide hints to narrow the search space, e.g. by coarsening the grid or reducing the range of values.

\subsection{Reducing Contention and Balancing Load}
\label{sec:macrobenchmark}

\begin{figure}[h]
\centering

\includegraphics[width=\columnwidth]{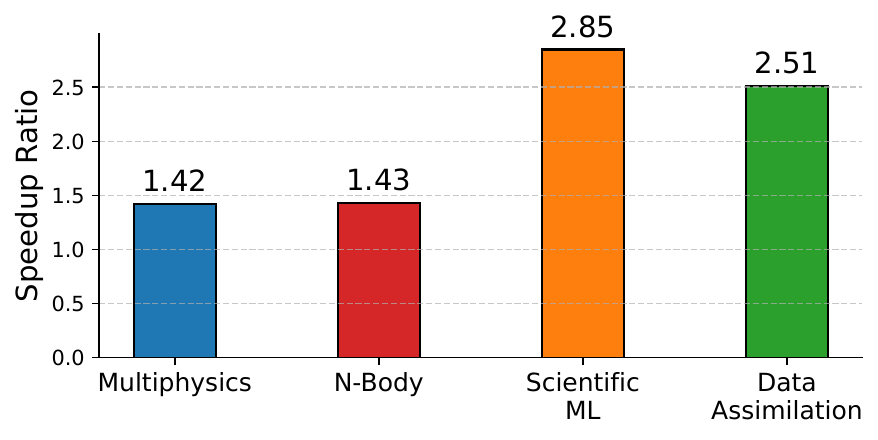}
\caption{Speedup achieved by using \VLCs on representative synthetic applications (higher is better).}\vspace{-5pt}
\label{fig:macrobenchmark}
\end{figure}

Due to the vast design and software codebases, we devised synthetic experiments based on the Rodinia 3.1 test suites \cite{rodinia}, \openblas kernels and \libtorch models  listed in \autoref{table:programmability}. We compose different benchmarks to resemble typical workflows in scientific computing.
\begin{enumerate} 
    \item {\bf Multiphysics workflow:} Two small-sized Hotspot3D (a heat transfer calculation) with a medium-sized CFD (computational fluid dynamics) and Cholesky (direct symmetric factorization): This workflow resembles a multiphysics/multifidelity simulation that involves a dense linear solve \cite{multiphysics}.
    
    \item {\bf N-body workflow:} Combinations of multiple GEMM, GESV, and Cholesky solves of different problem sizes that resemble a hierarchical matrix approximation factorization, a common technique for boundary integral equation solvers and Jacobians of neural networks \cite{nbody}.

    \item  {\bf Scientific Machine Learning Workflow:} Combination of a CFD code with  Kmeans and DNN, resembling a physics simulation that is corrected \& augmented with machine learning corrections \cite{scientific_ml}.

    \item {\bf Data assimilation workflow:}  Transformer + many small CFD resembling a time-series transformer followed by an ensemble CFD calculation \cite{data_assimilation}.
\end{enumerate}

We evaluate the performance improvements provided by \VLCs in eliminating resource contention (\textbf{RQ3}).
This involves comparing performance under two configurations: a baseline without \VLCs, and a setup where \VLCs partition the resources. In both cases, the libraries operate with their default configurations.
For each workflow, between 3 to 4 instances of \VLCs are created, each with a different CPU allocation. The ability to partition resources between libraries at a fine granularity also enables performance improvements through better load balancing. Here, we assigned more cores to the larger workloads in an experiment.

\autoref{fig:macrobenchmark} shows the speedup of composing benchmarks with \VLCs relative to composing without \VLCs. The result indicates that \VLCs can improve performance by up to 2.85$\times$, with an average speedup of 2.05$\times$ across all synthetic experiments, demonstrating that \VLCs
effectively reduce contention and improve load balancing by virtualizing and partitioning resources for libraries.

\subsection{Improving Nested Parallelism}

To evaluate \textbf{RQ4}, we consider workloads dependent on dense matrix multiplication. 

Libraries such as \openblas can only be configured with a single fixed level of parallelism 
by setting a process-level environment variable, making it impossible to tune the level of 
parallelism for each matrix size in the application. 
 
Programmers are forced to choose a best fit configuration, with the best choice often being to not overlap the execution of large and small GEMMs.

\begin{figure}[h]
\includegraphics[width=\columnwidth]{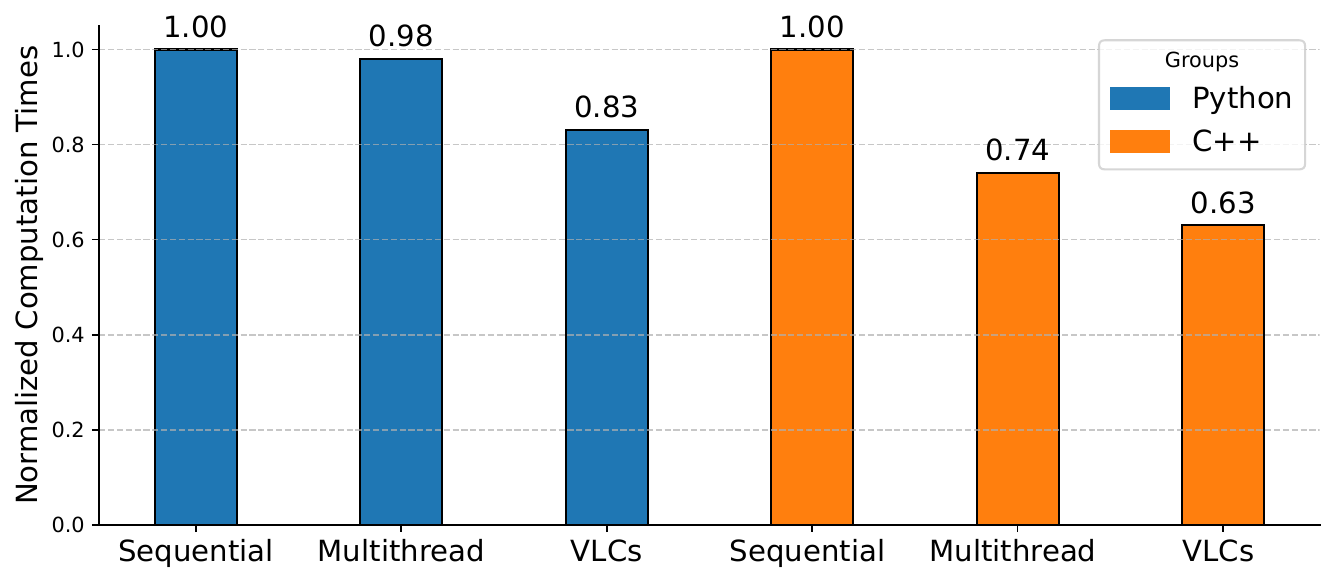}
\caption{Normalized execution time of matrix multiplication implementations (lower is better). Using \VLCs to manage nested parallelism yields an improvement of up to 1.58$\times$ over the sequential baseline. }
\label{fig:matmul}
\end{figure}

We explore such a scenario in which we perform 21 independent GEMMs of square matrices: 
twenty \(2048 \times 2048\) GEMMs and a single \(8192 \times 8192\) GEMM. Each 
GEMM is computed using the \openblas \texttt{cblas\_dgemm} routine in C++ and \texttt{numpy.matmul} 
in Python (numpy uses \openblas under the hood). \autoref{fig:matmul} shows the execution times for different implementations.

The baseline performs all GEMMs sequentially, each using all 24 cores in our machine. 
This allows the large matrix to execute fast, while consuming unnecessary resources for the small GEMMs. 
When parallelized with a conventional approach in C++, the multi-threaded (pthreads) implementation 
obtains a $1.3\times$ speedup. With Python multi-threading, the benefits of parallelization are negated 
by the overheads of the global interpreter lock for the short numpy calls of the small matrices. 
While reducing the core allocation of \openblas may allow multiple GEMMs to run in parallel, 
this hurts performance because reducing the number of cores for the large matrix so 
substantially degrades overall performance. 

With \VLCs, we can expose nested parallelism and execute the large and small GEMMs concurrently by dedicating most of the cores 
to the \VLC that performs the large GEMM and the rest of the cores to another \VLC that performs the small GEMMs. 
This approach works well, providing a $1.58\times$ speedup over the C++ sequential baseline ($1.17\times$ over the multithreaded version) 
and $1.2\times$ over the Python sequential baseline ($1.17\times$ over the multithreaded version). 
Allocating 17 cores to the large GEMM and 7 cores to the small GEMM results in the best performance. 

\subsection{Parallelizing Thread-Unsafe Calls}
We evaluate \textbf{RQ5} by exploring how 
\VLCs can be used to replicate libraries that are not thread-safe to enable safe concurrent calls to those libraries.

ARPACK~\cite{arpack} is a Fortran library that is utilized by popular 
libraries like SciPy~\cite{scipy} and MATLAB for solving sparse eigenvalue and singular value decomposition problems. 
Internally, ARPACK uses an unsynchronized static state, so simultaneous calls into ARPACK are not thread-safe \cite{scipy-arpack-threadsafe}. 
Libraries that depend on ARPACK have to guard all calls to ARPACK routines with a lock to avoid race conditions. 
This restriction can be relaxed by importing ARPACK into different \VLCs, where the separate static state in each instance makes simultaneous calls safe.

\begin{figure}[h]
\includegraphics[width=\columnwidth]{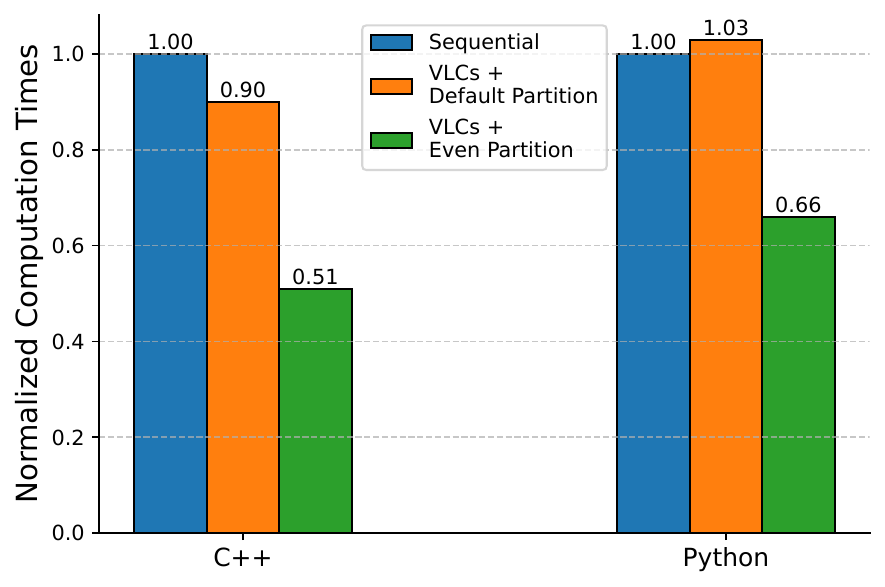}
\caption{Execution time for computing the eigenvalues of two matrices using ARPACK (lower is better).
The default partition loads two instances of ARPACK without specifying resource splits. The even partition allocates 12 cores to each ARPACK call. 
}\vspace{-5pt}
\label{fig:arpack}
\end{figure}

\autoref{fig:arpack} shows the results of computing the top 10 eigenvalues for two $10000\times 10000$ matrices using ARPACK.
Two instances of ARPACK are loaded into \VLCs with either no resource allocation specified or  allocating 12 cores to each. 
The Python implementations use {\tt scipy.sparse.linalg.eigsh} as a proxy to the ARPACK routine. 
The baseline serial execution without \VLCs does not utilize resources well. \VLCs offer up to a $1.96\times$ speedup in C++ ($1.51\times$ in Python) by tuning the cores allocated to each ARPACK call and running two calls concurrently.

\begin{figure}[h]
    \centering
    \includegraphics[width=\columnwidth]{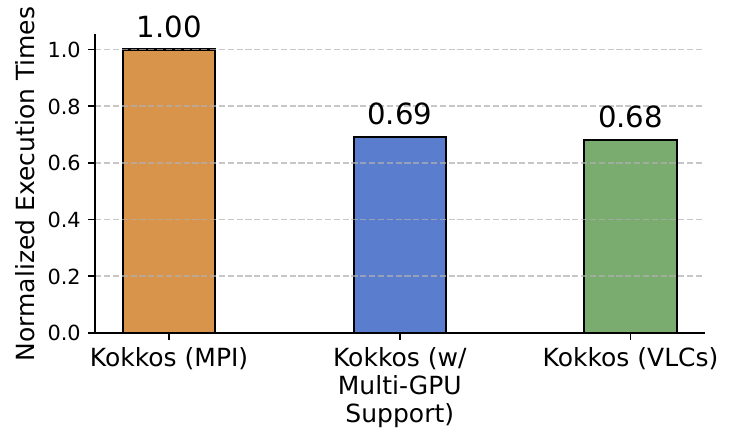}
    \caption{Normalized execution time of Kokkos multi-GPU Heat3D implementations (lower is better).
    Using \VLCs to manage multiple Kokkos instances yields a $1.46\times$ speedup over the MPI implementation and achieves identical performance to newer Kokkos native multi-GPU support.}
    \label{fig:kokkos}
\end{figure}

\subsection{Performance Comparison with MPI}
\label{sec:kokkos}
Kokkos~\cite{kokkos} has become a cornerstone C++ library for high-performance scientific computing.
It provides a unified programming model that abstracts the underlying hardware, allowing developers to write parallel programs that can be compiled across heterogeneous architectures.
Historically, Kokkos only supported a single GPU per process \cite{kokkos_gpu_issue}. Developers who wished to use Kokkos with multiple GPUs on a single machine needed to decompose their application into multiple processes and use IPC frameworks for data sharing, such as MPI. This multi-process MPI-based design has become the de facto standard for a vast ecosystem of Kokkos applications \cite{kokkos_mpi}.

While single-process multi-GPU support was recently added in Kokkos 4.3 (for CUDA) and Kokkos 4.6 (for HIP), it also brings a migration challenge. Porting an existing MPI-based Kokkos application requires code changes for data sharing and stream synchronization. Migrating a large body of legacy MPI-based codes to the new native multi-GPU model is a non-trivial task. \VLCs offer an alternative approach for enabling single process multi-GPU support for Kokkos by allowing multiple instances of Kokkos to be loaded into separate \VLCs, where each manages a separate GPU. This approach enables developers to continue using their existing MPI-based Kokkos applications without significant code changes.

We ported Kokkos's Heat3D MPI-based implementation ~\cite{kokkos_remote_space} to thread-based orchestration both using \VLCs and with the recent native multi-GPU support. We compare three implementations: (1) Kokkos with CUDA-aware MPI, (2) Kokkos with \VLCs, and (3) Kokkos with native multi-GPU support. Comparing these approaches allows us to evaluate the performance and migration cost of porting legacy Kokkos MPI-based applications to use multi-GPU in a single process.

This application solves the 3D heat equation on a rectangular box with dimensions \( L_x \times L_y \times L_z \) using a Forward Time-Centered Space finite-difference scheme. We solve the problem in a zero-temperature heat bath, with radiative heat loss on its surfaces, and an incoming heat flux on the bottom (\(z=0\)) surface that is removed halfway through the simulation.
In our experiment, the domain is split evenly between 2 GPUs, each holding \(L_x \times L_y \times \frac{L_z}{2}\) grid points.
At every time step, the GPUs exchange the \(L_x \times L_y\) boundary values between them. In the MPI implementation, CUDA-aware MPI is used to avoid copying data to the host and back when communicating between processes. 
We evaluate on a \(200 \times 200 \times 200 \) cube.

\autoref{fig:kokkos} shows the resulting performance of different Heat3D implementations. We can see that avoiding inter-process communication overheads when exchanging data between GPUs leads to a significant reduction in application runtime, even when avoiding host communication. Both the native multi-GPU support and support through \VLC isolation provide a $1.46\times$ speedup over the typical MPI implementation. \VLCs provide a competitive alternative that achieves the same performance as using the modern multi-GPU API, while working on legacy versions of Kokkos prior to 4.3. 

We note that in terms of migration costs the \VLC implementation required less direct code modification than porting the application to use multi-GPU execution spaces. 
The code within each \VLC is nearly identical to that within each MPI process, the only difference is that MPI send and receives are replaced with direct memory copies.  
In contrast, native multi-GPU support required more careful data management and changes to kernel launching parameters and CUDA stream synchronization throughout.
While no longer required to enable single process multi-GPU support in Kokkos, \VLCs provide a low-overhead within-application solution to overcome library configuration limitations that would typically require multiple processes.

\section{Discussion}
\subsection{Loading Incompatible Libraries}
\label{sec:incompatible}
\openblas, Atlas \cite{atlas}, and MKL \cite{mkl} are different implementations of the C-language BLAS standard. Applications are given the flexibility to choose an implementation based on its performance needs and environment. To reach optimal performance for various environments and application settings, a BLAS also provides many versions to choose from, such as different parallel backends, i.e., pthreads or \openmp.

However, when multiple libraries that use BLAS are composed together in a single process, they can only dynamically link a single BLAS library regardless of the fact that the different libraries may have their own preferences. It is also not possible to compose libraries that statically link to different BLAS builds as all BLAS implementations share the same symbols, which would cause name conflicts.

The incompatibility between different BLAS implementations can be addressed by using \VLCs. Applications can load different BLAS builds into multiple \VLCs, where each is isolated with linker namespaces so name conflicts will not occur. Each BLAS instance will also have private static data when loaded into a \VLC, making it safe to run in parallel with other BLAS libraries within a single process.

\subsection{Limitations}
Each \VLC creates a linker namespace to avoid name conflicts and provides private static data when loading multiple instances of the same shared library into a single process. The maximum number of \VLCs is limited by the number of linker namespaces provided by the dynamic linker. The dynamic linker provides tunable \texttt{glibc.rtld.nns} \cite{glibc-tunables} to set the limit of linker namespaces up to 16. While 16 \VLCs should be sufficient for most applications, it is possible to modify glibc to increase the limit to 32 or more.

\subsection{Future Work}
The current implementation of \VLCs focuses on virtualizing CPU cores and GPU devices, but the underlying principle of library-level resource virtualization holds significant potential for broader applicability. A primary direction for our future research is to explore the extension of this paradigm to other critical system resources, including the virtualization of memory capacity and network interfaces, which could enable more fine-grained control over application performance in distributed and memory-bound scenarios. 

While many applications achieve optimal performance through straightforward partitioning schemes (e.g., uniform distribution or workload-based allocation), these approaches are often insufficient for complex hardware topologies. For these non-trivial scenarios, such as optimizing resource allocation across multi-socket NUMA architectures, an adaptive tuning mechanism is essential. Our current auto-tuner employs a grid search policy, and this exhaustive approach can be computationally expensive. To address this, our future efforts will focus on building a machine learning model that can intelligently prune the search space, enabling the auto-tuner to converge on near-optimal configurations within a significantly reduced number of iterations.

\section{Conclusion}
We explore \LLVfullname with \VLCs, subunits of a process that encapsulate sets of libraries and provide performance isolation between them. \VLCs enable applications to partition compute resources among libraries, avoiding contention while retaining the benefits of a shared address space. Evaluations highlight how \VLCs improve performance by eliminating resource contention, balancing workloads, safely parallelizing otherwise thread-unsafe code, and enabling nested parallelism that is not natively supported.

\begin{acks}
We thank our anonymous reviewers for their helpful comments, which strengthened our work. This work was partially funded by the U.S. Department of Energy, National Nuclear Security Administration Award Number DE-NA0003969; by the NSF CISE “Expedition” Grant Number 2326576; and by NSF award SHF-2505085.
\end{acks}

\bibliographystyle{ACM-Reference-Format}
\bibliography{ref}

\end{document}